\documentclass[conference]{IEEEtran}
\IEEEoverridecommandlockouts
\usepackage{algorithmic}
\usepackage{textcomp}
\usepackage{xcolor}
\def\BibTeX{{\rm B\kern-.05em{\sc i\kern-.025em b}\kern-.08em
    T\kern-.1667em\lower.7ex\hbox{E}\kern-.125emX}}

\usepackage{cite}      
\usepackage[cmex10]{amsmath}
\usepackage{amssymb}
\usepackage{amsfonts}
\usepackage{mdwlist}
\usepackage{ifpdf}
\ifpdf
\usepackage[pdftex]{graphicx}
\graphicspath{{./pics/}}
\DeclareGraphicsExtensions{.pdf}
\else
\usepackage[dvips]{graphicx}
\graphicspath{{./pics/eps}}
\DeclareGraphicsExtensions{.eps}
\usepackage{breakurl}
\fi
\usepackage[caption=false,font=footnotesize]{subfig}
\usepackage{upgreek}
\usepackage[colorlinks=true,linkcolor=blue,citecolor=blue,urlcolor=blue]{hyperref}
\usepackage{pinlabel}
\usepackage[scaled]{helvet}

\usepackage{booktabs}
\usepackage{multirow}

\usepackage[flushleft]{threeparttable}
\usepackage{multicol}
\usepackage{balance}

\begin{document}

\title{Random Convolution Kernels with Multi-Scale Decomposition for Preterm EEG Inter-burst Detection

\thanks{This research was supported by Science Foundation Ireland (15/SIRG/3580).}
}

\author{
\IEEEauthorblockN{Christopher Lundy$^{1,2}$ and John M. O'Toole$^{1,2}$}
\IEEEauthorblockA{
\textit{$^{1}$Irish Centre for Maternal and Child Health Research (INFANT), University College Cork, Ireland.} \\
\textit{$^{2}$Department of Paediatrics and Child Health, University College Cork,
    Ireland. } \\
\textit{christopher.lundy@ucc.ie, jotoole@ucc.ie}} }

\maketitle

\begin{abstract}

  Linear classifiers with random convolution kernels are computationally efficient methods that need no design or domain knowledge.  Unlike deep neural networks, there is no need to hand-craft a network architecture; the kernels are randomly generated and only the linear classifier needs training.  A recently proposed method, RandOm Convolutional KErnel Transforms (ROCKETs), has shown high accuracy across a range of time-series data sets.  Here we propose a multi-scale version of this method, using both high- and low-frequency components.  We apply our methods to inter-burst detection in a cohort of preterm EEG recorded from 36 neonates $<$30 weeks gestational age.  Two features from the convolution of 10,000 random kernels are combined using ridge regression.  The proposed multi-scale ROCKET method out-performs the method without scale: median (interquartile range, IQR) Matthews correlation coefficient (MCC) of 0.859 (0.815 to 0.874) for multi-scale versus 0.841 (0.807 to 0.865) without scale, $p<0.001$.  The proposed method lags behind an existing feature-based machine learning method developed with deep domain knowledge, but is fast to train and can quickly set an initial baseline threshold of performance for generic and biomedical time-series classification.
\end{abstract}


\section{Introduction}
\label{sec:introduction}
    Deep learning generalises across a variety of applications with particular efficacy when applied to large quality data sets \cite{Fawaz2019}.  The self learning of modern deep neural networks (DNNs) has produced state-of-the-art classification performances without the need for user-specific domain knowledge \cite{Lecun2015}.  Machine learning using \emph{hand-crafted features} is often critiqued in deep learning studies due to the reliance on domain knowledge and time consuming selection or design of features \cite{Miotto2017}.  In task-specific implementations however, machine learning requires skilled feature analysis and DNN design requires a prolonged trial-and-error process (network \emph{crafting}).  The effort of designing and training a DNN, however, has frequently out-performed machine learning for large data sets \cite{Lecun2015}.  Even in studies with fewer data points, such as neonatal electroencephalogram (EEG) applications, DNNs can provide a performance that justifies the effort \cite{Raurale2021,Ansari2020}.
    

    Here we seek a different approach: can we apply a machine-learning system, with relatively good performance, that is neither designed nor optimised for a specific application?
    This zero-knowledge system would provide a baseline performance on which all other application-specific methods should improve on. It should also be fast and easy to implement.
    
    We take advantage of recent studies developing and comparing general-purpose algorithms applied to a wide variety of time-series classification problems  \cite{Fawaz2019,Dempster2020,Anh2018}.  Many different approaches have been developed as generic classifiers, with varying levels of computational efficiency.  Deep learning models, for example, are inherently adaptable to these generic problems \cite{Fawaz2019}, however they can be cumbersome to train and difficult to settle on a constant network architecture. 
    A recently proposed method using random convolutional kernels, known as ROCKET (RandOm Convolutional KErnel Transform), can be trained in a small fraction of the time compared to DNNs, with only a slight reduction in accuracy \cite{Dempster2020}.

    The ROCKET method consists of a very wide single layer of random kernels connected to a linear classifier.  The production of many kernels (tens of thousands) can match patterns of complex shape and frequency in generic time-series data.  
    Two features of each convolution with the kernel are fed forward to the linear classifier: the maximum value and the proportion of positive values (PPV).  
    The PPV adds an element of nonlinearity, similar in some ways to the rectified linear unit in DNNs.  
    Without any parameters to select or optimise, ROCKET is a zero-knowledge machine learning method that is fast and easy to train.
    It can set a baseline performance level for time-series data such as EEG, in a \emph{one-size-fits-all} type model.
    
    \begin{figure*}[ht!]
      \centering
      \includegraphics[scale=0.45]{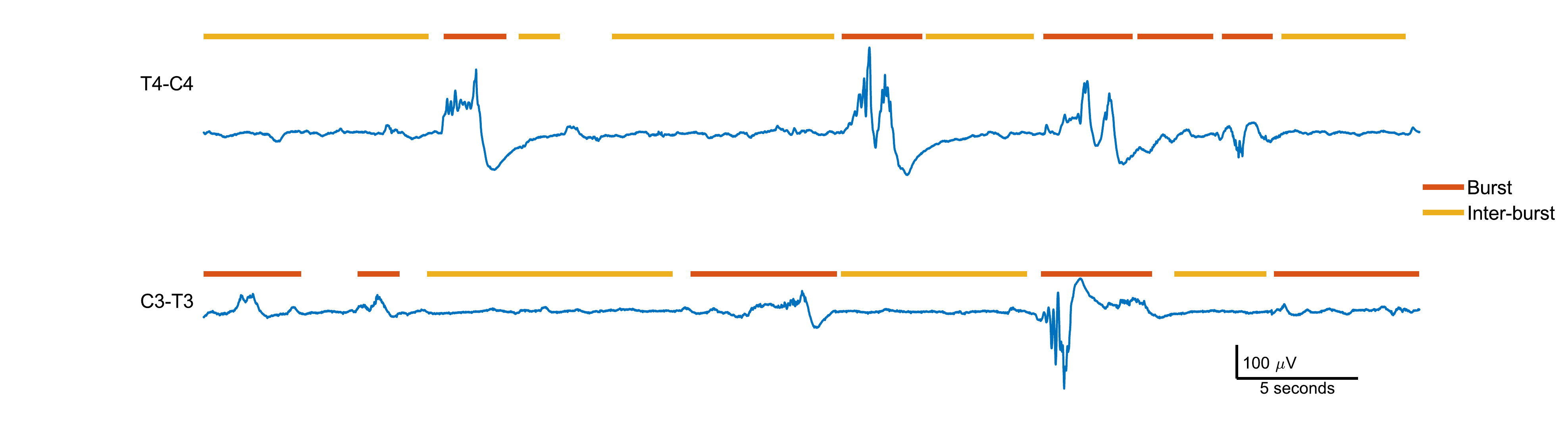}
      \caption{Annotations of bursts and inter-bursts of a 50-second EEG segment for 1-channel from 2 different preterm infants.}
      \label{fig:eeg_anno_example}
    \end{figure*}    
    
    
    The ROCKET method includes a dilation parameter, which downsamples the time-series data before convolution. 
    To avoid aliasing, a probable outcome when downsampling without an aliasing filter, we first filter the time-series data. 
    This results in a multi-scale decomposition of the signal. We also extract the high-pass components from the decomposition, in addition to the default low-pass components generated by this multi-scale analysis. The low-pass component is downsampled.
    We compare detection performance of different aspects of the multi-scale ROCKET method when applied to the problem of detecting inter-bursts in preterm EEG.
    Inter-burst waveforms are characteristic of preterm EEG and are important markers for estimating brain maturity and neurological well being.



\section{Materials and Methods}
\label{sec:methods}
\subsection{Preterm EEG data set}
    Individual EEG recordings from $36$ preterm infants were acquired within $72$ hours of birth in the NICU of the Cork University Maternity Hospital, Ireland.  Gestational age ranged from $23.4$ to $29.7$ weeks with a median age of $27.5$ weeks.  EEGs were acquired using the NicoletOne EEG system (Natus Medical Incorporated, USA) with a minimum sampling rate of $256$ Hz.  Eleven electrodes were used in a modified version of the international 10--20 system, with a reference electrode at Fz and a ground electrode behind the left ear.  Babies with severe brain injuries were excluded from analysis; brain injury was determined by a cranial ultrasound within the first week post-birth.  Informed and written consent was obtained before EEG recording.  Data collection was approved by the Cork Research Ethics Committee of Cork Teaching Hospitals, Ireland.
    
    One channel was analysed for each EEG recording, alternating over the set of bipolar channels F4--C4, C4--O2, T4--C4, C4--Cz, F3--C3, C3--O1, C3--T3 and Cz--C3.  A continuous segment per EEG was annotated separately by two clinical physiologists.  This segment was on average $10$-minutes in duration (range: $8.2$ to $10.2$ hours), had minimal artefact, and was taken at a mean of $14$-hours (range: $3$ to $41$ hours) after-birth.  This data set is further detailed in O’Toole \emph{et al.} \cite{OToole2017}.  Examples of inter-burst annotations are shown in Fig.~\ref{fig:eeg_anno_example}.
    
\subsection{ROCKET implementation}
    The ROCKET method convolves a large number ($10,000$) of random kernels with the EEG, extracting 2 features from each convolution operation.  The random kernels contrasts to the training process of DNNs; however, the variety and number of kernels can match elements of complex patterns in the EEG and other time-domain data.  The features extracted from the convolution, representative of different time-domain waveforms, are combined in a linear classifier using ridge regression \cite{Dempster2020}.  

    We express the convolution as the following correlation operation
    \begin{equation}
      \label{eq:1}
      x[n] * k[n] = \sum_{m=0}^{M-1}k[m]x[n+md]
    \end{equation}
    for epoch $x[n]$ of length-$N$ and kernel $k[n]$ of length-$M$.  Zero-padding of $x[n]$ avoids wrap-around effects from the correlation. 
    Because the kernel $k[n]$ is a random sequence we use the terms convolution and correlation inter-changeably throughout.  The degree of padding is a random parameter of the method.  Dilation factor $d$, another random parameter of the method, sub-samples the epoch $x[n]$ when $d>1$.  Because no anti-aliasing filtering is present, this will likely result in aliasing for $x[n]$ and will not capture patterns at different scales.

    We propose a few changes to this approach.  First, we eliminate padding as a random parameter of the method and always include sufficient padding to avoid the wrap around effects.  We do so to eliminate the role of padding from further analysis.  Thus, $x[n]$ is extended from length $N$ to length $N+M-1$ by padding zeros at the start and end of $x[n]$.  Second, we include scale $s$ as a random parameter,
    generated from the exponential distribution
    \begin{equation}
      \label{eq:2}
      s = \lfloor 2^a \rfloor,\quad a \sim \mathcal{U}(0, \log_2\{1 + \lfloor N / M \rfloor\})
    \end{equation}
    where $\mathcal{U}$ is the uniform distribution. When $s>1$, we first perform a moving-average filtering of $x[n]$ before the convolution operation in \eqref{eq:1}:
    \begin{equation}
      \label{eq:3}
      y_{\textrm{low}}[n - s_h] = \frac{1}{s} \sum_{q=0}^{s-1} x[n-q]
    \end{equation}
    for $n=s_h, s_h + 1, \ldots, N + s_h -1 $, with shift $s_h = \lfloor s/2 \rfloor$.  This shift centres the moving-average windowing and results in a length-$N$ low-pass filtered signal $y_{\textrm{low}}[n]$.  We next define the high-pass component as the residual of $x[n]$ minus the low-pass component:
    \begin{equation}
      \label{eq:4}
      y_{\textrm{high}}[n] = x[n] - y_{\textrm{low}}[n]
    \end{equation}
    A random parameter, a binary variable with equal probability, determines whether to include the high-pass or low-pass component in the convolution.  For the low-pass signal $y_{\textrm{low}}[n]$, we include a dilation parameter by downsampling $y_{\textrm{low}}[ns]$, where $s$ is the scale value.

    Our convolution operation is then as follows
    \begin{equation}
      \label{eq:5}
      y[n] * k[n] = \sum_{m=0}^{M-1}k[m]\hat{y}[n-ms]
    \end{equation}
    where $\hat{y}[n]$ is the zero-padded version of $y[n]$ to avoid wrap around artefacts and $y[n]$ is either $y_{\textrm{low}}[n]$ or $y_{\textrm{high}}[n]$.  Thus, our kernels have 5 random parameters:
    \begin{enumerate}
        \item kernel length $M$, drawn with equal probability from the set $\{7,9,11\}$ \cite{Dempster2020} 
        \item zero-mean $k[n] \sim \mathcal{N}(0,1)$ of length-$M$, where $\mathcal{N}$ is the normal distribution \cite{Dempster2020}
        \item bias value $b \sim \mathcal{U}(-1,1)$, where $\mathcal{U}$ is the uniform distribution \cite{Dempster2020}
        \item scale $s$, defined in (\ref{eq:2})  
        \item binary variable (yes/no) to decide whether to use the high- or low-frequency component.  
    \end{enumerate}
        
    The moving-averaging operation in (\ref{eq:3}) and residual calculation in (\ref{eq:4}) will incur extra computation over the original definition.  As there are many (10,000) kernels per epoch $x[n]$, this computational load can be minimised with caching of the high- and low-frequency components for each scale.

\begin{figure}[h!]
      \centering
      \includegraphics[scale=0.55]{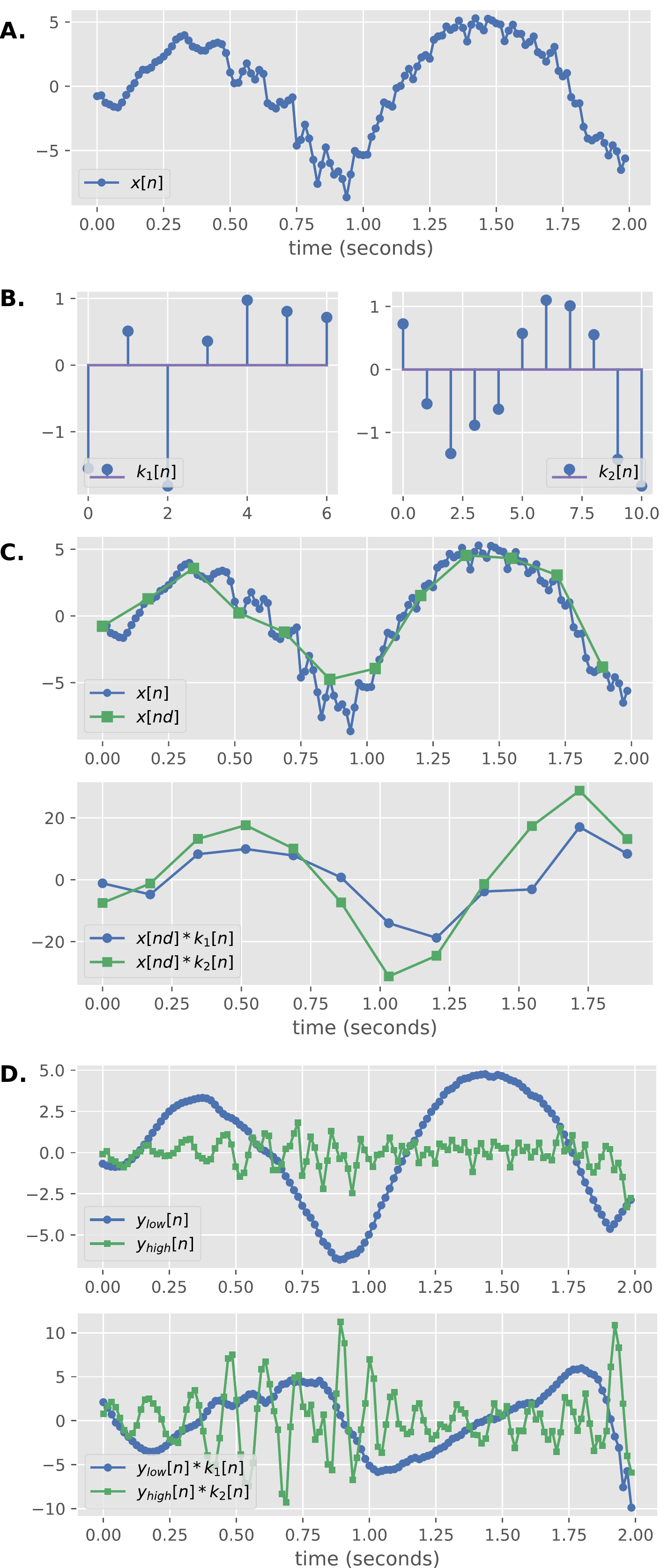}
      \caption{Convolution example.  A: Normalised 2-second EEG epoch $x[n]$;  B: Kernels
        $k_1[n]$ of length-7 and $k_2[n]$ of length-11. C: EEG epoch $x[n]$ with
        downsampled epoch $x[nd]$ for dilation $d=11$ (top); convolution output from
        downsampled component $x[nd]$ with kernel $k_1[n]$ and kernel $k_2[n]$ (bottom).
        D: Low-frequency component $y_{\textrm{low}}[n]$
        generated from an 11-sample moving-average window on $x[n]$ and high-pass component as the
        residual of signal and low-pass component ($y_{\textrm{high}}[n] = x[n] - y_{\textrm{low}}[n]$)
        (top); convolution outputs from low-pass component and kernel $k_1[n]$ and
        high-pass component and kernel $k_2[n]$ (bottom).}
      \label{fig:kern_conv}
    \end{figure}

    Fig.~\ref{fig:kern_conv} presents an illustration of the convolution process for an EEG epoch with 2 kernels. The figure illustrates the process for the original ROCKET method which downsamples the EEG with dilation factor $d$ (Fig.~\ref{fig:kern_conv}C). The decomposition of the EEG epoch into high- and low-frequency components for the proposed method and subsequent convolution with the kernels are shown in Fig.~\ref{fig:kern_conv}D.  
    %

    Two features are extracted for each kernel after the convolution operation, thus resulting in a total of 20,000 features per EEG epoch. These features are:
    \begin{itemize}
    \item Maximum value of the convolution operation 
    \[
    \textrm{MAX} = \max\{ x[n] * k[n]\}
    \]
    \item Proportion of positive values
    \[
    \textrm{PPV} = \frac{1}{N} \sum_{m=0}^{N-1}[(x[n] * k[n])_m +b >0]
    \]
    \end{itemize}
    Where $b$ is the bias scalar and $[\cdot]$ represents the Iverson bracket, that is $[P]=1$ when $P$ is true and $[P]=0$ otherwise.


    To assess the utility of the modifications to the convolution kernel from the original
    ROCKET method, we try 4 different experiments on the preterm EEG data set:
    \begin{enumerate}
    \item No scaling (set $s=1$): $x[n] * k[n] + b$ 
    \item Multi-scaling: $y_{\textrm{low}}[n] * k[n] + b$ \space when $s>1$
    \item Multi-scaling with high- and low-frequencies:  \newline
        $y_{\textrm{high}}[n] * k[n] + b$ \space or \space 
        $y_{\textrm{low}}[n] * k[n] + b$  \space when $s>1$
    \item Multi-scaling with high- and low-frequencies\newline and dilation:\newline 
      $y_{\textrm{high}}[n] * k[n] + b$  \space or \space $y_{\textrm{low}}[ns] *k[n] +b$ \space  when $s>1$
    \end{enumerate}

    The multi-scale ROCKET methods are available as Python code at \url{https://github.com/otoolej/ms_rocket} (version 0.1).

\subsection{Training and testing}
    EEG was downsampled from 256 Hz to 64 Hz after applying an anti-aliasing filter \cite{OToole2017} and then normalised by subtracting the median and dividing by the interquartile range (IQR). 
    Normalisation is a requirement of the ROCKET method \cite{Dempster2020}.  
    Median and IQR estimated during training were used in testing.
    EEG was segmented into $2$-second epochs with with an overlap of $25\%$ for training and $93.75\%$ for testing. Overlap was lower for training, with an equivalent sampling frequency of $2$ Hz, to reduce computational load. To better assess performance, overlap for testing was higher with an equivalent sampling frequency of $8$ Hz, above and beyond the resolution of the experts' annotations.
    In training, a burst or inter-burst label was assigned if either exceeded an arbitrary threshold of greater than $90\%$ of samples per epoch.
    Reducing this threshold below 100\% allowed us to include more data for training, in particular epochs closer the start or end of inter-bursts.

    Using a leave-one-out cross-validation loop we iterate through the left-out preterm EEG creating a testing and training set with one and $35$ preterm neonate(s) per loop respectively.  A ridge regression classifier is trained and tested on the $20,000$ features ($2$ features per convolution of $10,000$ kernels) for each $2$-second epoch.  The $\ell_2$ regularisation parameter $\alpha$ was set to the the default value $\alpha=1$. 

    Performance was analysed by Matthews correlation coefficient (MCC), a more descriptive summary measure of the confusion matrix for imbalanced data sets than the commonly used (in neonatal EEG literature \cite{Palmu2010b,Koolen2014,OToole2017,OToole2019}) area under the receiver operator characteristic curve (AUC).  The proposed ROCKET inter-burst detector methods are compared with an existing multi-feature support vector machine (SVM) method \cite{OToole2017}.  We compare the differences in MCC 
    using the Wilcoxon signed-ranked test.

\section{Results}

\label{sec:results}
\begin{figure}[h!]
  \centering
  \includegraphics[scale=0.62]{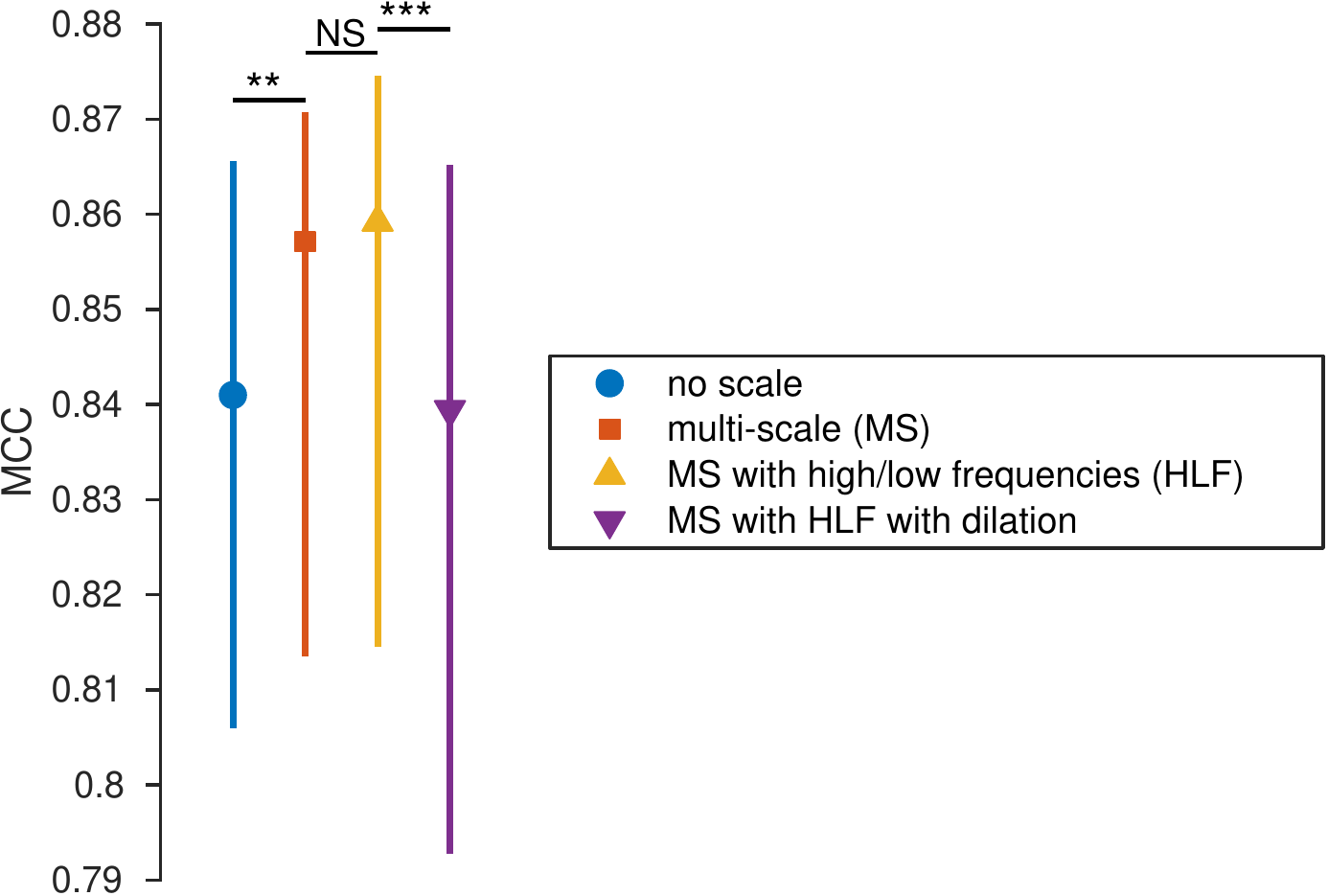}
  \caption{Performance for inter-burst detector in preterm EEG using random convolutional
    kernels.  Matthews correlation coefficient (MCC) from leave-one-baby out testing
    results for the 4 different approaches to kernel convolutional.  Significance: $**$ for
    $P<0.01$; ${*}{*}{*}$ for $P<0.001$; and NS for not significant ($P>0.05$) using the Wilcoxon signed-ranked test.  Dots, squares, and triangles represent median values; lines
    represent the interquartile (IQR) range.}
  \label{fig:MCC_results}
\end{figure}

\begin{table}
    \begin{threeparttable}
    \centering
    \caption{Comparison of inter-burst detection methods using different random kernel approaches.  Also included is the
    existing multi-feature machine learning method \cite{OToole2017}.}
    \begin{tabular}{llll}
        \toprule
        & MCC & MCC \\
        model                       & median (IQR)           & range\\
        \midrule
        Random kernel methods: & & \\
        No scale                           & 0.841 (0.807 to 0.865) & (0.512 to 0.913)\\
        Multi-scale (MS)                   & 0.857 (0.814 to 0.871) & (0.503 to 0.916)\\
        MS-HLF                             & 0.859 (0.815 to 0.874) & (0.495 to 0.916)\\
        MS-HLF and dilation                & 0.839 (0.798 to 0.865) & (0.476 to 0.918)\\
        \\
        Knowledge-based approach: & & \\
        Multi-feature SVM                 & 0.866 (0.813 to 0.897) & (0.668 to 0.925)\\
        \bottomrule
    \end{tabular}
        \begin{tablenotes}
        \small
        \item\footnotesize{Key: IQR, interquartile range; HLF, high- and low-frequencies; MS, Multi-scale; SVM, support vector machine; MCC, Matthews correlation coefficient}
        \end{tablenotes}
    \end{threeparttable}
    \label{tab:tab_results}
\end{table}
    Detection performance for the methods are presented in Table~\ref{tab:tab_results} and illustrated in Fig.~\ref{fig:MCC_results}.
    The multi-scale (MS) ROCKET model with a median (IQR) MCC of 0.857 (0.814 to 0.871) was higher than the method without scale 0.841 (0.807 to 0.865), $p < 0.01$.
    The MS high- and low-frequencies (MS-HLF) model achieved best performance of the implemented ROCKET methods with median (IQR) MCC of 0.859 (0.815 to 0.874).  The MS-HLF significantly ($p<0.001$) outperformed the MS-HLF with dilation.
    There was no significant difference between the MS-HLF and MS model, however both significantly ($p < 0.01$) out-performed the standard no scale ROCKET model, see Fig.~\ref{fig:MCC_results}.

    The MS-HLF ROCKET model was able to detect inter-bursts with a high level of accuracy. 
    Nonetheless, as expected, all the general-purpose ROCKET models under-performed the application-specific multi-feature SVM model \cite{OToole2017}.  The difference in performance was small however: median (IQR) MCC difference of 0.007 (0.002 to 0.023).
    
Computational time for the proposed methods is slightly longer than the original ROCKET method \cite{Dempster2020}, see Table~\ref{tab:tab_comp_results} for a comparison.  This table summarises computational time for the moving-average filtering and kernel convolutions for 10,000 kernels with an epoch length of 128 samples.  These tests were performed on a desktop computer with a 6 core CPU (Intel i7-8700K CPU at 3.7 GHz) and 32 GB of RAM.  Convolutions were implemented in parallel.  The convolution operations account for approximately two-thirds of the total training time; the other one-third is for training the ridge regression with the 20,000 features per epoch.\newline

\begin{table}
    \begin{threeparttable}
    \centering
    \caption{Computation time in seconds for convolution operations of 10,000 random kernels with $n$-epochs of 128 samples.}
  \begin{tabular}{llll}
    \toprule
                                                  & $n=1,000$  & $n=10,000$  \\
                                                  & mean (SD)  & mean (SD)   \\
    \midrule                                  
    ROCKET method \cite{Dempster2020}             & 1.3 (0.01) & 13.4 (0.05) \\
    No scale                                      & 1.8 (0.02) & 17.9 (0.28) \\ 
    Multi-scale (MS)                              & 1.8 (0.02) & 18.8 (0.48) \\ 
    MS with high- and low-frequencies (HLF)       & 1.9 (0.00) & 19.7 (0.12) \\ 
    MS-HLF and dilation                          & 1.5 (0.02) & 14.8 (0.16) \\
    \bottomrule
  \end{tabular}
  \begin{tablenotes}
    \small
  \item\footnotesize{Key: HLF, high- and low-frequencies; MS, Multi-scale; SD, standard deviation;}
  \end{tablenotes}
\end{threeparttable}
\label{tab:tab_comp_results}
\end{table}

\section{Discussion}
\balance
    Convolving the random kernels at multiple scales has a significant performance improvement ($p < 0.001$) in our application of EEG inter-burst detection.  This improvement in performance could be because the proposed method overcomes the negative effects of aliasing caused by dilation in the original ROCKET method.  Both the MS ROCKET with and without the high- and low-frequency divisions significantly outperformed the method without the filtering.  The best performing model included both the high- and low-frequencies (MS-HLF), which may in some ways replicate the decomposition of the EEG into the standard frequency bands.  These processes are not directly comparable however, as the multi-scale filtering decomposes into high- and low-frequency components only, whereas the filtering operation used for EEG decomposes into discrete bands.

    The general-purpose ROCKET model with MS-HLF was able to detect inter-bursts with a high level of accuracy in preterm EEG, but it did not perform as well as the multi-feature SVM model \cite{OToole2017}, as detailed in Table~\ref{tab:tab_results}.  The difference in performance is small however, with a median (IQR) MCC decrease of 0.007 (0.002-0.023) between the two methods.  This is not a surprising result as the feature-based approach was designed and implemented with years of learned domain knowledge for the specific task of detecting inter-bursts in the preterm EEG \cite{OToole2017}.  Nor does it negate the utility of the ROCKET approach---it requires no domain knowledge, needs no design, and is fast to train and test.
 
The ROCKET method produces 2 features from the convolution operation.  The MAX feature is similar to the global-max pooling measure in CNNs, a common measure to reduce temporal or spatial variance and dimensionality.  There is an important difference here however: there is no nonlinear function after the convolution operation as there are in CNNs.  The PPV on the other hand, does implement a nonlinear function by setting negative values to zero, similar to the rectified linear unit. The PPV is a proportional measure of the positive values produced when the EEG matches samples of a given random pattern. Bias, $b$, acts as a threshold for a stronger pattern matching measure when negative, and includes weaker matches when positive. In the original ROCKET method, PPV was found to be a more important feature than MAX, and both together proved best \cite{Dempster2020}.

Our results show that a multi-scale anti-aliasing approach does improve on the basic ROCKET method, for our specific application. In future work we could compare performance with a broader array of time-series classification problems, for example by using the UCR archive \cite{Dempster2020,Anh2018}.  Future work could also include a band-pass decomposition to replace the multi-scale decomposition, by using a discrete wavelet transform.  Although this transform would be computationally efficient compared to a filter-bank approach, our multi-scale ROCKET uses the residual of a low-frequency signal to obtain a high-frequency component and is therefore not equivalent to band-pass filtering.  The MS-HLF ROCKET also omits the dilation factor that generates downsampling but it is a part of the discrete wavelet decomposition.

 The multi-scale ROCKET model could be used as a fast and easy implementation to set a baseline performance in future neonatal EEG applications.  We suspect in some applications that a baseline performance will suffice, and certainly an important first step for comparison with bespoke solutions. The method is applied without prior knowledge of the signal characteristics or without the need to design a feature set or neural network architecture.  A one-size-fits-all \emph{press play} model for time-series classification and regression problems.


\begin{thebibliography}{10}
\providecommand{\url}[1]{#1}
\csname url@samestyle\endcsname
\providecommand{\newblock}{\relax}
\providecommand{\bibinfo}[2]{#2}
\providecommand{\BIBentrySTDinterwordspacing}{\spaceskip=0pt\relax}
\providecommand{\BIBentryALTinterwordstretchfactor}{4}
\providecommand{\BIBentryALTinterwordspacing}{\spaceskip=\fontdimen2\font plus
\BIBentryALTinterwordstretchfactor\fontdimen3\font minus
  \fontdimen4\font\relax}
\providecommand{\BIBforeignlanguage}[2]{{%
\expandafter\ifx\csname l@#1\endcsname\relax
\typeout{** WARNING: IEEEtran.bst: No hyphenation pattern has been}%
\typeout{** loaded for the language `#1'. Using the pattern for}%
\typeout{** the default language instead.}%
\else
\language=\csname l@#1\endcsname
\fi
#2}}
\providecommand{\BIBdecl}{\relax}
\BIBdecl

\bibitem{Fawaz2019}
H. ~I. ~Fawaz, H., G. ~Forestier, J. ~Weber, L. ~Idoumghar, P. ~Muller, ``{Deep learning for time series classification: a review},'' \emph{Data Mining and Knowledge Discovery}, vol.~33, pp. 917--963, 2019.

\bibitem{Lecun2015}
Y.~Lecun, Y.~Bengio, and G.~Hinton, ``{Deep learning},'' \emph{Nature}, vol. 521, no. 7553, pp. 436--444, 2015.

\bibitem{Miotto2017}
R.~Miotto, F.~Wang, S.~Wang, X.~Jiang, and J.~T. Dudley, ``{Deep learning for healthcare: review, opportunities and challenges},'' \emph{Brief. Bioinform.}, vol.~19, no.~6, pp. 1236--1246, 2017.

\bibitem{Raurale2021}
S.~A. Raurale, G.~B. Boylan, S.~R. Mathieson, W.~P. Marnane, G.~Lightbody, and J.~M. O’Toole, ``{Grading hypoxic-ischemic encephalopathy in neonatal EEG with convolutional neural networks and quadratic time–frequency distributions.}'' \emph{J. Neural Eng.}, vol.~18, no.~4,  2021

\bibitem{Ansari2020}
A.~H. Ansari, O.~De Wel, K.~Pillay, A.~Dereymaeker, K.~Jansen, S.~{Van Huffel}, and M.~{De Vos}, ``{A convolutional neural network outperforming state-of-the-art sleep staging algorithms for both preterm and term infants.}'' \emph{J. Neural Eng.}, vol.~17, no.~1, 2020.

\bibitem{Dempster2020}
A.~Dempster, F.~Petitjean, and G.~Webb, ``{ROCKET: exceptionally fast and accurate time series classification using random convolutional kernels},'' \emph{Data Mining and Knowledge Discovery}, vol.~34, pp. 1454--1495, 2020.

\bibitem{Anh2018}
H.~Dau, E.~Keogh, ~Kamgar, ~Kaveh, C.~Yeh, Y.~Zhu, S.~Gharghabi, A.~Ratanamahatana, ~Yanping, B.~AndHu, N.~Begum, A.~Bagnall, A.~Mueen, G.~Batista, ~Hexagon-ML, ``{The UCR Time Series Classification Archive},'' 2018.

\bibitem{OToole2017}
J.~M. O'Toole, G.~B. Boylan, R.~O. Lloyd, R.~M. Goulding, S.~Vanhatalo, and N.~J. Stevenson, ``{Detecting bursts in the EEG of very and extremely premature infants using a multi-feature approach},'' \emph{Med. Eng. Phys.}, vol.~45, pp. 42--50, 2017.

\bibitem{Pavlidis2017b}
E.~Pavlidis, R.~O. Lloyd, S.~Mathieson, and G.~B. Boylan, ``{A review of important EEG features for the assessment of brain maturation in premature infants},'' \emph{Acta Paediatr.}, vol.~38, no.~1, pp. 42--49, 2017.

\bibitem{OToole2016b}
J.~M. O'Toole, G.~B. Boylan, S.~Vanhatalo, and N.~J. Stevenson, ``{Estimating functional brain maturity in very and extremely preterm neonates using automated analysis of the electroencephalogram},'' \emph{Clin. Neurophysiol.}, vol. 127, no.~8, pp. 2910--2918, 2016.

\bibitem{Palmu2010b}
K.~Palmu, N.~Stevenson, S.~Wikstr{\"{o}}m, L.~Hellstr{\"{o}}m-Westas, S.~Vanhatalo, and J.~M. Palva, ``{Optimization of an NLEO-based algorithm for automated detection of spontaneous activity transients in early preterm EEG.}'' \emph{Physiol. Meas.}, vol.~31, no.~11, pp. N85--93, 2010.

\bibitem{Koolen2014}
N.~Koolen, K.~Jansen, J.~Vervisch, V.~Matic, M.~{De Vos}, G.~Naulaers, and S.~{Van Huffel}, ``{Line length as a robust method to detect high-activity events: automated burst detection in premature EEG recordings.}'' \emph{Clin. Neurophysiol.}, vol. 125, no.~10, pp. 1985--94, 2014.

\bibitem{OToole2019}
J.~M. O'Toole, G.~B. Boylan, ``{Machine learning without a feature set for detecting bursts in the EEG of preterm infants},'' \emph{41st Annual International Conference of the IEEE Engineering in Medicine \& Biology Society (EMBC)}, pp. 5799--5802, 2019.

\end{thebibliography}
\end{document}